# The Canada-France Redshift Survey II: Spectroscopic program; data for the 0000-00 and 1000+25 fields

Olivier Le Fèvre[1]
DAEC, Observatoire de Paris-Meudon, 92195 Meudon, France

David Crampton[1]
Dominion Astrophysical Observatory, National Research Council of Canada, Victoria, Canada

Simon J. Lilly[1]
Department of Astronomy, University of Toronto, Toronto, Canada M5S 1A7

Francois Hammer[1] and Laurence Tresse[1]
DAEC, Observatoire de Paris-Meudon, 92195 Meudon, France

## ABSTRACT

This paper describes the methods used to obtain the spectroscopic data and construct redshift catalogs for the Canada-France deep Redshift Survey (CFRS). The full data set consists of more than one thousand spectra, of objects with $17.5 \leq I_{AB} \leq 22.5$, obtained from deep multi-slit data with the MARLIN and MOS-SIS spectrographs at the CFHT. The final spectroscopic catalog contains 200 stars, 591 galaxies with secure redshifts in the range $0 \leq z \leq 1.3$, 6 QSOs, and 146 objects with very uncertain or unknown redshifts, leading to an overall success rate of identification of 85%. Additionally, 67 objects affected by observational problems have been placed in a supplemental list.

We describe here the instrumental set up, and the observing procedures used to efficiently gather this large data set. New optimal ways of packing spectra on the detector to significantly increase the multiplexing gain offered by multi-slit spectroscopy are described. Dedicated data reduction procedures have been developed under the IRAF environment to allow for fast and accurate processing.

Very strict procedures have been followed to establish a reliable list of final spectroscopic measurements. Fully independent processing of the data has been carried out by three members of the team for each data set associated with a multi-slit mask, and final redshifts were assigned only after the careful comparison of the three independent measurements. A confidence class scheme was established. We strongly emphasize the benefits of such procedures.

Finally, we present the spectroscopic data obtained for 303 objects in the 0000-00 and 1000+25 fields. The success rate in spectroscopic identification is 83% for the 0000-00 field and 84% for the 1000+25 field.

---

[1]Visiting Astronomer, Canada-France-Hawaii Telescope, which is operated by the National Research Council of Canada, the Centre de Recherche Scientifique of France and the University of Hawaii

*Subject headings:* galaxies: distances and redshifts – galaxies: evolution – Methods: data analysis

## 1. INTRODUCTION

Multi-slit spectroscopy is an extremely powerful tool which can provide spectra with S/N identical to traditional single long-slit spectrographs, but, with the same amount of integration time, also yield a large number of spectra. This, in effect, considerably increases the effective light gathering power of the telescope being used, and opens new fields of investigations.

One of the research areas benefiting most directly from the large multiplexing power of multi-object spectroscopy is the study of faint field galaxies. To date, large amounts of observing time have been used to measure redshifts of relatively small samples of faint galaxies. The deepest redshift surveys so far have been conducted on galaxies selected in the B and K bands (Broadhurst at al. 1988; Colless et al. 1990, 1993; Cowie et al. 1995; Songaila et al. 1995). Altogether, the current versions of these surveys provide redshift measurements for only a few tens of galaxies above redshifts of a half.

A sample large enough to investigate the evolution of the properties of galaxies up to z∼1 with a good confidence level had yet to be defined. Two deep I-band selected surveys were started at the CFHT with the MARLIN spectrographs and results on several tens of redshifts were presented by Lilly (1993) and Tresse et al. (1993). The commissioning of the MOS-SIS imaging spectrograph at CFHT in 1992 (Crampton et al. 1992; Le Fèvre et al. 1994) offered a combination of a wide field, a large area CCD and spectral resolution which allows ∼100 objects to be observed simultaneously, significantly better than achieved at most other facilities. This allowed more ambitious deep survey programs to be contemplated and prompted the two teams to merge their efforts and to undertake the Canada France Redshift Survey. The goal of obtaining spectra for 1000 objects selected solely from their brightness in the range $17.5 \leq I_{AB} \leq 22.5$ from 5 independent survey fields was adopted. We present here the observing strategy and techniques used to bring this goal to fruition.

This paper is the second paper in a series of 5 papers presenting the CFRS core observations. CFRS-I (Lilly et al., 1995) describes the overall science goals and the methods used to prepare unbiased photometric catalogs from which targets were selected for spectroscopy; this paper presents the strategy of multi-slit spectra acquisition, data reduction, and the methods used to construct the spectroscopic catalogs for the whole survey, and gives the spectroscopic data for the 0000-00 and 1000+25 fields; CFRS-III discusses single emission line objects and the statistics of repeat observations and presents data for the 1415+52 and 2200+00 fields (Lilly et al., 1995); CFRS-IV examines possible biases in the spectroscopic data and presents data for the 0300+00 field (Hammer et al., 1995); CFRS-V applies basic checks to the photometric and spectroscopic data and considers the properties of the sample as a whole (Crampton et al., 1995). Scientific analyses will be presented in subsequent papers.

This paper is organized as follows: §2 describes the strategy adopted for the acquisition and calibration of multislit spectra in a minimum amount of time; §3 describes the data reduction techniques and 3-fold data processing that lead to the redshift measurements and the spectroscopic catalogs; §4 presents the spectroscopic data for 303 redshifts in the 0000-00 and 1000+25 fields of the CFRS.

## 2. DATA ACQUISITION WITH MULTI-OBJECT SPECTROSCOPY

### 2.1. Preliminary multi-slit spectroscopy with the MARLIN spectrograph

The initial surveys were carried out with the MARLIN spectrograph at the CFHT and have been described by Lilly (1993) and Tresse et al. (1993). A total of 94 objects have been observed with MARLIN and these observations have helped determine the observing parameters adopted for the CFRS.

Tresse et al. compared the redshift completeness achieved with two different spectral resolutions, $17\text{\AA}$ and $35\text{\AA}$, and concluded that these provided indistinguishable results. This analysis had important implications in determining the observing strategy for subsequent deep redshift surveys, since the lower spectral resolution allows simultaneous observations of 3 times more galaxies with the MOS-SIS spectrograph as is described in the next section.

The minimum size of the slits to be used was also determined from these observations. Since most galaxies towards the faint magnitude limit of the CFRS have isophotal radii of 2-3 arcseconds, a slit 12 arcsecond in length would contain sky background information on more than 75% of its length. With a sampling of ~0.3 arcsec per pixel, or 40 pixels along the slit, about 30 pixels would then be used to fit low order polynomials for the sky background subtraction, which we find provided adequate results in terms of sky background residuals.

The MARLIN observations were eventually included in the CFRS catalogs as described in §3.2.

### 2.2. The CFHT/MOS and the "multi-layer spectra" concept: "strip" geometry and object selection

The MOS-SIS imaging-spectrograph came into operation in mid-1992 (Crampton et al., 1992; Le Fèvre et al., 1994). In the Multi-Object-Spectograph (MOS) mode, this instrument is coupled to a laser cutting machine to use aperture masks with slits cut to high precision in a $9.'4 \times 8.'3$ field. Detailed descriptions of the instrument and the observing procedures are given in Crampton et al. (1992) and Le Fèvre et al. (1994); detailed operational procedures can be found in Le Fèvre (1994).

The optimization of the multiplexing gain is the result of two factors: the equivalent slit length projected on the sky, and the length of one spectrum projected on the detector. In the low resolution mode of the MOS spectrograph with the V150 grism, one spectrum from $4250\text{Å}$ to $8500\text{Å}$ covers 590 pixels on a $2048^2$ CCD with 15 $\mu$m pixels, hence one can in principle easily fit 3 spectra on the CCD along the dispersion direction, with about 90 arcseconds left as latitude to select the targets. This lead to the "3 strip concept" demonstrated in Figure 1: objects were selected for each mask in 3 narrow bands each about $30''\text{x}9.'4$, separated by $210''$. In the spatial direction, the instrument field is 9.4 arcmin, hence the total slit length on the sky is about 28 arcmin, which in theory allows fitting in 140 adjacent slits 12 arcseconds in length each. The optimum projected density of objects to take advantage of the full multiplexing gain would then be on order 36000 galaxies per square degree. For $I_{AB} \leq 22.5$ galaxies, the projected sky density is $\sim$20000 galaxies per square degree, so that in practice, owing to spectral and spatial overlap constraints, we were able to fit an average of 70 to 80 slits per mask. Note that we sometimes had more than one object per slit.

To fit this geometry on the detector, we restricted the wavelength range of spectra with respect to the total spectrograph range. We used a highly efficient custom interference filter with $\sim$95% transmission from $4250\text{Å}$ to $8500\text{Å}$. This wavelength range was selected because it allows measurement of features in the rest-frame region 3700-4350 Å, which is central to most redshift measurements, from z=0.1 to z$\sim$1 (see CFRS-IV).

This strategy implies two limitations on the spectra: (i) the spectrum from a slit above another slit in the dispersion direction will be merged with the second order of the spectrum of the bottom slit, (ii) the zero order of an upper slit (and included object) will be projected on the spectrum of an object directly below it. Tests conducted prior to the start of the survey showed that both order-overlapping effects would have minimal impact on our ability to measure redshifts. Figure 2 shows the relative strength of the first order and second order spectra for the V150 grism which was used throughout the survey; the second order is never more than 1.5% in intensity relative to the first order. Figure 3 shows the intensity distribution in the dispersion direction through a projected zero order; the maximum extent of the zero order is about 150Å, and the core is about 3 times brighter than the 5577Å sky line. Standard sky line subtraction techniques can frequently be used to cleanly subtract all but the brightest core (Figure 3).

Objects were selected for spectroscopy from the photometric catalog (see CFRS-I) in the range $17.5 \leq I_{AB} \leq 22.5$ solely on the basis of position. The target position was chosen from a diagram in which objects were represented as points (to completely avoid any personal selection bias), so as to allow 3 layers of spectra to fit on the CCD with no overlap. The length of slits was adjusted in most cases in order to have the overlapping zero order completely cover the slit length of the spectrum being overlapped to allow for easier order subtraction. We deliberately avoided discrimination against compact objects: althougth this category contains a large fraction of stars, it also contains compact galaxies and QSOs. We investigate in CFRS-V the consequences that such selection would have implied.

The selection of objects in subsequent masks of a same field was done in the same strips in which the previous observations were done. This allowed observation of close pairs that would otherwise be missed in a single mask setting because of the minimum length of slits. Moreover, this strategy allowed us also to re-observe objects on which we had failed during previous observations.

The spatial distribution of the galaxies observed is strongly dependent on the strip geometry adopted. Figure 4 shows the ratio of the number of pairs for the galaxies observed in spectroscopy over the number of possible pairs in the photometric catalogs vs. the pair separation. One can easily identify "bumps" in this distribution at $\sim 210''$ and $\sim 420''$ which correspond to the average inter-strip angular separation. However, on scales smaller than $\sim 60''$, or about the average resulting width of strips explored in spectroscopy after $\sim 3$ masks, this distribution shows that our data is a representative sample of the overall projected pair separation.

## 2.3. Spectroscopic observations

We have used slits $1\rlap{.}''75$ in width and the V150 grism. This led to a spectral resolution of $40\mathring{A}$, as confirmed by the measurement of the width of arc calibration lines. Slits were oriented East-West on the sky to minimize the effects of differential atmospheric refraction as one goes to larger airmasses.

The steps that lead to efficient observations for this survey are as follows. A mask design was first produced ahead of the observing run based on the deep field photometric catalogs, and the brightness range and geometric constraints described in §2.3. The mask designs were done solely in term of position, without reference to any other object properties such as size, brightness, compactness or colors. At the start of a spectroscopic observing run, a 900 second direct image was obtained through MOS in the I filter. Objects in the mask design were then located on the MOS image and the mask was fabricated with the laser machine (15 min cutting time needed for $\sim 80$ slits). One or two reference holes located on bright stars (mag <19) were added in the mask to allow for accurate positioning of the mask and to allow monitoring of drift and re-positioning of the telescope between exposures if necessary. By viewing the reference star(s) through the reference hole(s) in direct imaging mode (10 sec exposure), the telescope position was adjusted to position the mask on the objects, with the help of precise telescope offset tools (accuracy $\sim 0.1$ arcsec). MOS was then configured in the spectroscopic mode and the first spectroscopic exposure taken (typically one hour). A quick check of the centering of the reference stars was performed between each exposure.

Observations with MOS were conducted in a series of observing runs indicated in Table 1.

The Lick2, Loral2 and Loral3, $2048^2$, $15\mu$m pixels, thick blue coated CCDs were used as they became available. In the few instances when absorption by cirrus was present we estimated in real time the average absorption for a given exposure by comparing the total flux for a star to the flux in a photometric exposure. Additional exposures were then taken to compensate for any loss and bring the equivalent exposure time close to 8 hours. The first observing run (October 1992) was conducted with a CCD with a poor blue coating leading to difficulties in flat fielding. Most of the unidentified objects in these October 1992 masks were re-observed in August 1993 and November 1993, only the mask 3 for 0300+00 could not be properly re-observed and this mask was abandoned. A total of 19 masks were observed over 21 nights. Of these, 19 were clear, and this program indeed benefitted from the very good Mauna Kea weather.

We have routinely achieved in excess of 8 hours of spectroscopic integration per night on our survey fields, broken down into 8 one hour exposures.

An average of three masks have been observed per field, with an effective covering factor of $\sim$22.4% per field. The total equivalent area surveyed by the whole CFRS is $\sim$112 arcmin$^2$. As shown in Figure 5, the CFRS spectroscopic sample is an unbiased sample from the photometric sample, as the faint I counts associated with the spectroscopic data are an excellent representation of the photometric counts after scaling by the covering factor.

### 2.4. Calibrations

Complete sets of calibrations were obtained for each mask. These include direct images of the masks (Figure 1a) to obtain the position of the slits projected on the CCD, Helium and Argon arc-lamp spectra, and flat-field spectra. In addition, observations of spectrophotometric standard stars (Oke 1974; Stone 1987) were obtained to flux calibrate the spectra.

Fits of the arc lines with low order splines give typical errors in the wavelength solution of $\sim 0.7$Å. Relative spectrophotometry across the spectra is generally accurate to 10% as indicated by comparison with the photometric data (see section 3.5). In some cases, the comparison of photometry with a spectrum indicates that the spectrophotometric accuracy over the full spectroscopic range can be significantly worse than this number.

### 3. DATA REDUCTION

### 3.1. MULTIRED package

The volume of data acquired for this survey required efficient data reduction tools, consequently, MULTIRED was developed as a general purpose package to process multi-slit spectra based on the "onedspec" and "twodspec" packages available in IRAF. In principle, the processing of multi-slit spectra differs from long-slit processing only by the number of slits to process, and the careful accounting that needs to be performed. After the creation of a file with positions of the slits projected on the CCD, MULTIRED processes each of the slits in turn, and manages all of the corresponding calibration information necessary.

Careful preparation of the data set is necessary before using MULTIRED. First, a distortion correction is performed to correct for the geometric distortion induced by the spectrograph optics. A CCD image of an aperture mask with a grid of points of known spacing cut with the laser machine was used to generate a correction map by fitting the distortion with 2D cubic splines, and each image was corrected from this map (IRAF tasks geomap/geotran). Then, each of the several (usually 8) spectral images was checked and corrected for possible shifts in the location of spectra from one frame to the next, introduced by flexures between the spectrograph and the offset guider. Shifts are generally found to be less than a pixel (in which case no correction is applied), but in a few occasions when observations of a same mask were spread over several nights, shifts of 1 to 2 pixels had to be applied.

MULTIRED is then used to perform the following steps in sequence for each slit:

- Extraction of the 2D spectra of the object, and the corresponding wavelength calibration and flat field, from the full $2048^2$ pixels images.

- For each 2D spectrum (8 per mask/per slit on average):

    - Correct for flat field (pixel to pixel variations).
    - Subtract sky emission: the sky is fitted with adjustable low-order polynomials and subtracted along the slit for each wavelength element (Figure 6). A treatment of the zero-order superposition was also added: areas on the 2D spectra with a zero order could be corrected independently from the rest of the spectrum if needed.

- Combine all of the corrected 2D spectra, with either an average or median scheme using sigma-clipping rejection. (With $\sim$ 8 2D spectra to combine, this removes most of the cosmic ray events, although in some circumstances, the brightest events can still partly remain).

- Extraction of a 1D spectrum of the arc lamps and cross-correlation with a reference arc lamp spectrum to produce an initial wavelength solution. The fit was then adjusted if necessary. This produces a unique pixel/wavelength transformation for each slit.

- Extraction of a 1D spectrum from the corrected 2D spectrum for each object of interest in the slit.

- Wavelength and flux calibration of the 1D spectrum.

- Plot the corrected and calibrated 1D spectrum, and display all of the corrected 2D spectra together with the averaged corrected 2D spectrum (Figure 7). Line identification for redshift measurement can then proceed, assessing the reality of a line on the 1D spectrum by comparison with the multiple 2D spectra.

The accuracy of the data reduction process is illustrated in Figure 8 where we plot the residuals left after the data reduction, including the sky subtraction, expressed as a percentage of the sky brightness. Between 4250 and 8500$\text{Å}$, the r.m.s. residuals are 0.25% of the sky brightness, with a remarquably good correction even in the reddest part of the wavelength range.

A typical mask with 80 slits can be fully processed in about 2 working days with MULTIRED using Sun/Sparc-10 class machines, including a mix of interactive and batch processes.

### 3.2. Independent data reduction

The complexity of the data reduction and the notorious difficulty of measuring redshifts of high redshift galaxies, for which spectral properties might differ significantly from the local population, makes deep redshift surveys especially sensitive to personal biases, which might lead to significant statistical errors. This prompted us to adopt a scheme by which, for each mask, three individuals fully and independently reduced the spectroscopic data, identified the spectrum and assigned a confidence class to that identification. This was done without any knowledge of the photometric or morphological properties of the object. Although based on the same raw data, the data reduction follows a decision-tree where choices of data reduction parameters can result in slight but sometimes significant changes in the final spectrum. During the reduction process, the software allows multiple choices of the key parameters that were left to each person's judgment to adjust. Of particular importance are the parameters selected for spectroscopic flat fielding, sky background removal, the method of combining the 2D spectra, and the choice of the extraction window used to create the 1D spectrum. A classic example was the possibility to combine the individual 2D spectra after flat field and sky correction by using a median scheme or an average with sigma-clipping. In the first case, cosmic ray events are efficiently discarded, while the sigma-clipping average sometimes left traces of the brightest cosmic ray hits which can mimic emission lines. On the other hand, the sigma-clipping average sometimes provides better continuum and feature information, and the two choices may then bring complementary information.

After the individual data reductions, a face to face meeting of the three team members assigned to a mask reduction (see Table 1) took place. A comparison of the three spectra sets was conducted to assign a confidence notation on the redshift measurements for all measurements that agreed, and open a (sometimes lengthy) debate for each object for which one or more of the individual identifications disagreed. This debate was based on the recognised necessity for all to agree on spectral features claimed to support the measured redshift, and was conducted with constant reference to the 2D and 1D data. In some instances, complete new processing of a spectrum was obtained to verify, on both the 2D and 1D spectra, the validity of a feature identified by a team member. When a redshift measurement achieved a consensus from all three team members, the redshift and a confidence notation were assigned to the object and one of the spectra was chosen to best represent the final redshift. All identified features in the spectrum agreed upon by all were noted. In those instances where no redshift measurement would satisfy all three members, the object was classified as unidentified. It should be stressed that this comparison process was again conducted without reference to morphological or photometric properties of the objects.

Each object identification was thus assigned a confidence class. This scheme is somewhat different from a pure "quality" scheme (e.g. based on the S/N of the spectra), and the confidence class reflects the consensus probability that a redshift measurement is correct and represents the confidence in the measurement one gets from a combination of S/N, the number and agreement of features, and the continuum shape. The notation was set to classes 0 to 4, 8, 9, 12 to 14, 92 to 94 as follows. Confidence class 1 was meant to imply a probability of 50% that the measurement was correct. Class 2 was meant to be more than 75% correct. These were for identifications that appear to be relatively secure, with a moderate S/N, and/or small number of matching lines. Class 3 was set to be at least 95% secure with a good number of matching lines and supportive continuum information, class 4 was reserved for unquestionably correct identifications. When no redshift could be assigned, a class 0 was assigned to the object and the redshift written as "9.9999" in the spectroscopic catalog. Single emission line objects were put in a separate class and were examined in detail at a later stage in the survey. These are objects for which the spectrum shows only one secure emission line which leads to an ambiguity in the line identification and hence in the redshift, if e.g., the line was to be assigned to [OII]3727 or H$\alpha$. An algorithm based on the shape of the continuum was devised to resolve this ambiguity (see CFRS-III), and led to the classification of these objects in either class 8 or 9. Class 8 is for objects for which the algorithm indicates that the emission line is [OII]3727, class 9 is for objects for which the redshift ambiguity still could not be resolved. QSOs are identified with the same quality notation 1-4 as galaxies, but a 1 is placed in front, e.g. 14 is a very secure QSO. Objects which do not belong to the main catalog, either because they have $I_{AB}$ >22.5 or $I_{AB}$ <17.5, have been rejected because of instrumental problems (see §3.3) but have a redshift determination, or whose photometry was adjusted fainter than $I_{AB}$ =22.5 after the spectroscopic observations (see CFRS-I), are kept in a supplemental catalog, and identified by a 9 in front of the confidence class, e.g., 93. The objects in this supplemental catalogue may thus have biases that will not be present in the statistically complete sample.

In the case of objects observed more than once, we processed each spectrum independently as above. The individual spectra were co-added in some instances, and the final redshift and a confidence class assigned.

These repeat observations allow us to establish an empirical calibration of the confidence classes (see CFRS-III). Through comparison of the observations, we are able to directly assign a level of confidence for each of the classes. These objective empirical tests confirm our original subjective estimates of the reliability of our confidence classes to a very high degree: they give confidence levels of 50% (class 1), 85% (class2 ), 95% (class 3), 100% (class 4).

The spectra of objects coming from the early MARLIN observations were not reduced in triplicate in the same way that the MOS spectra were. Nevertheless, each 1D MARLIN spectra was reviewed by the whole team and assigned a confidence class.

The process of comparing spectra processed differently by 3 team members, although time consuming, was found extremely useful in detecting incorrect identifications, in estimating the reliability of the claimed identification, and in enhancing the final quality of the spectra. We found this process to be particularly healthy.

### 3.3. Identification of instrumental problems

We were particularly careful to set strict rules to decide whether an object spectrum should be rejected from the final catalog based on purely instrumental reasons. For example, the brightest objects, or those with strong emission lines, even when affected by instrumental problems, might yield a secure redshift while the faintest may not and the inclusion of these objects in the catalogue would have led to statistical biases. The instrumental reasons that led us to reject spectra from the main catalog are the following:

- A defective CCD column along the dispersion direction, less than 5 pixels from the center of the object.

- Large and abrupt variations of the sky brightness along the slit of more than 5% (due to a bad cut by laser, excessive dust, bad flat fielding).

- Less than 2 pixels to properly estimate the sky background on one side of the spectrum (edge of the slit placed too close to the object).

- The projection of zero order(s) from slit(s) above implying a loss of more than $500\text{Å}$.

- The projection of the second order from slits below causing a major contamination (case of the brightest $I_{AB} \sim$17-18 stars only).

- Object not in slit. A position check was performed between the object position in the photometric catalog and the slit position in the masks. Any object with a discrepency between these positions of more than 1.5 arcsecond perpendicular to the slit length was rejected. In this context, it has to be noted that a spectrum was never rejected simply because nothing was seen in the slit, as long as the positional check showed no discrepancy.

The objects that were rejected this way, but were nevertheless identified, were kept in a supplementary catalog, as described in §3.4. The others were eliminated.

### 3.4. Redshift accuracy

From a technical point of view, the redshift measurement accuracy is limited by the following main factors: (i) the accuracy of the slit positioning relative to the object, (ii) the accuracy of the wavelength calibration, (iii) the accuracy in the line measurement. The wavelength calibration fit gives on average a residual of $\sim 0.7 \text{\AA}$ r.m.s, and the accuracy of line position measurements is on order $1 \text{\AA}$ (1/7th of a dispersion element), but these two factors are minor contributors compared to the slit positioning error. Errors in slit position are the result of three steps: (1) the mask design, for which one has to accurately estimate the brightness peak of each object before locating a slit with software for later laser cutting, this is done to an accuracy of about 0.2 arcsec; (2) the slit cutting which is done to a few microns accuracy and therefore contributes very little; and (3) mask positioning on the sky, which is done with an accuracy of 0.2 arcsec for each integration, but increases to $\sim 0.3$ arcsec after 8 integrations are combined. There is thus a total slit positioning error of $\sim 0.4$ arcsec which translates into a $8 \text{\AA}$ error in the zero point of the wavelength scale, or to $\sim 350$ km s$^{-1}$.

From duplicate spectroscopic observations the empirical redshift measurement error can be derived (CFRS-III). The r.m.s. redshift difference is 0.0026, or a velocity uncertainty per observation of $\sim 550$ km s$^{-1}$ when taking all 50 duplicate measurements for which the confidence class was 2 or greater. However, discarding the 5 duplicate measurements for which the redshift difference is larger than 0.005, indicates a r.m.s. redshift difference of 0.0018 or a velocity uncertainty per observation of $\sim 380$ km s$^{-1}$, comparable to the value predicted above.

### 3.5. Spectroscopic catalogs

Spectroscopic catalogs were compiled for each field, and incorporate the final redshift assignment, confidence class and spectral features retained after the 3-fold data reduction. Examples of spectra in each class are presented in Figure 9. Objects in the classes 2, 3, 4, 8, 9, 12, 13, 14 are considered as secure identifications, while objects in the classes 0 and 1 are considered as unidentified. Objects with classes 9* are placed in a supplementary catalog.

A total of 1010 objects have been observed spectroscopically of which 67 are in the supplementary catalog. In the statisticaly complete sample of 943 objects, 200 (21%) are secure stars, 591 (63%) are galaxies with secure redshifts, 6 (1%) are secure QSOs, 146 (15%) are unidentified objects. The distribution in classes is 330 objects in class 4 (35%), 313 in class 3 (33%), 29 (3%) in class 8, 33 (3%) in class 9, 92 (10%) in class 2, 53 (6%) in class 1, and 93 (10%) in class 0.

We present in section 4 the spectroscopic catalogs for the 0000−00 and 1000+25 fields. Spectroscopic catalogs for the 1415+52 and 2200+00 fields are presented in CFRS-III (Lilly et al.) and the 0300+00 field catalog in CFRS-IV (Hammer et al.).

### 4. SPECTROSCOPY IN THE 0000−00 AND 1000+25 FIELDS

### 4.1. Field identification

The coordinates for the center of the fields are:
$\alpha_{2000}=00^h\ 02^m\ 39\overset{s}{.}6$, $\delta_{2000}$ =-00°41′45″
$\alpha_{2000}=10^h\ 00^m\ 43\overset{s}{.}7$, $\delta_{2000}$=+25°14′05″
Grey scale plots of the 0000−00 and 1000+25 fields are provided in Figure 10.

### 4.2. Catalog of spectroscopic observations for the 0000−00 and 1000+25 fields

Two masks have been observed in the 0000−00 field leading to 90 spectra. This low yield is due to the large number of objects that had to be re-observed because they were unidentified in the first mask due to a deficient blue coating of the CCD in 1992 October, they were subsequently re-observed in 1993 August. The catalog includes 16 secure stars, 52 secure galaxies, 1 QSO, 14 unidentified objects, and 8 in the supplementary catalog. The identification rate in this field is therefore 83%. Table 2 lists the full spectroscopic catalog. Figure 11a identifies all the measured galaxies with their redshift among all the galaxies with $I_{AB}$ ≤22.5.

Three masks have been observed in the 1000+25 field in 1992 December and 1993 February, leading to 210 spectra. The catalog includes 27 secure stars, 143 secure galaxies, 33 unidentified objects and 7 in the supplementary catalog. The identification rate in this field is therefore 83.7%. Table 3 lists the full spectroscopic catalog. Figure 11b identifies all the measured galaxies with their redshift among all the galaxies with $I_{AB}$ ≤22.5.

The redshift histograms for these two fields are shown in Figure 12.

The format of Table 2 and Table 3 is as follows:

Column (1): The object identification number. The first two digits indicate the survey field, e.g., 00 indicates the 0000−00 field. The last four digits indicate the object number in the photometric catalog for that field.

Column (2): Right ascension in hours, minutes and seconds of time. Equinox is 2000. The relative accuracy in the coordinates is on the order of $0\overset{''}{.}2$, while the absolute accuracy is on order $1''$.

Column (3): Declination in degrees, arcminutes and arcseconds. Equinox is 2000.

Column (4): Isophotal magnitude in the I band expressed in the AB system ($I_{AB}$=I+0.48), measured in the $\mu_{I_{AB}}$=28 mag arcsec$^{-2}$ isophote (see CFRS-I).

Column (5): $(V-I)_{AB}$ color index measured in a 3 arcsec aperture.

Column (6): Q compactness parameter (see CFRS-I).

Column (7): Redshift. No corrections to transform to heliocentric or galactocentric systems were applied.

– 13 –

Column (8): Confidence class.

Columns (9+): Identified spectral features.

## 5. SUMMARY

We have presented the strategy and methods developed to maximize the observing efficiency and obtain the spectroscopic data for the Canada-France Redshift Survey of field galaxies with $17.5 \leq I_{AB} \leq 22.5$. High multiplexing gains have been achieved through multi-slit observations with the MOS spectrograph at CFHT, with an average of 80 slits per aperture mask. New data reduction tools used for the extraction of the spectra have been developed. These methods have allowed us to obtain more than a thousand spectra of objects in the redshift range $0 \leq z \leq 1.3$ in less than 19 clear nights with the CFHT.

Strict procedures have been set to process the spectroscopic data and establish the final spectroscopic catalogs. Classes have been set and calibrated to assign confidence levels to the measurements. Independent data reduction by three team members and a three-fold comparison of the measurements has provided a list of identified and unidentified objects, and has ensured a strict control of possible personal biases and instrumental effects. At the end of this process, the CFRS spectroscopic catalogs contain 1010 measurements of which 943 are in a statistically complete catalog with an identification rate of 85%.

Finally, we have presented the spectroscopic catalogs for 303 objects in the 0000−00 and 1000+25 survey fields.

We thank the CFHT directors and our respective time allocation committees for their continuous support to this project. SJL research is supported by NSERC. We acknowledge travel support from NATO, and support for publication from CFHT.

Table 1: Summary of Observations

| Date of observations | Fields observed | Mask | Exp. time (sec.) | Processed by |
|---|---|---|---|---|
| 1992 Oct 22-29 | 0000−01 | 1 | 27000 | FH-OLF-SL |
| | 0300+00 | 1 | 30000 (some cirrus) | DC-FH-OLF |
| | 0300+00 | 2 | 32300 (some cirrus) | DC-OLF-SL |
| | 0300+00 | 3 | 32100 (some cirrus) | Bad quality |
| | 2215+00 | 1 | 35600 (some cirrus) | DC-OLF-SL |
| | 2215+00 | 2 | 28800 | DC-FH-SL |
| 1992 Dec 28-29 | 0300+00 | 4 | 30000 (some cirrus) | DC-FH-OLF |
| | 0958+25 | 1 | 27300 | FH-OLF-SL |
| 1993 Feb 18-25 | 0300+00 | 5 | 24900 | DC-FH-SL |
| | 0958+25 | 2 | 28800 | DC-FH-OLF |
| | 0958+25 | 3 | 27900 | DC-OLF-SL |
| | 1415+52 | 1 | 34500 (some cirrus) | DC-FH-SL-LT |
| | 1415+52 | 2 | 27900 | FH-OLF-SL |
| | 1415+52 | 3 | 18000 | DC-FH-OLF |
| 1993 May 15 | 1415+52 | 3 | 4500 | DC-FH-OLF |
| 1993 Aug 15-18 | 0000−01 | 2 | 28800 | FH-OLF-SL-LT |
| | 0300+00 | 6 | 33000 | DC-FH-OLF |
| | 2215+00 | 3 | 28800 | DC-OLF-SL |
| 1993 Nov 9-10 | 0300+00 | 7 | 28000 | FH-OLF-SL |
| | 0300+00 | 8 | 31200 | DC-OLF-SL |

Fig. 1.— *(a)* Image of a mask as designed from the photometric catalog. *(b)* A 2D image of the spectra obtained with CFHT/MOS through the mask.

Fig. 2.— The spectrum of a flat-field continuum lamp demonstrating that the relative strength of the second order with the V150 grism is at the most 1.5% of the first order intensity.

Fig. 3.— *(a)* Left: a raw 2D spectrum before correction; the zero order has an intensity (dominated by sky emission) about 3 times stronger than the sky [OI]5575Å line. Right the flat field, sky-corrected, and zero-order subtracted 2D spectrum. *(b)* Same as (a) but cut along the wavelength direction. After subtraction of the zero-order done after fitting with a low order polynomial, the affected wavelength range is not more than 160Å where most of the information under the zero order is lost.

Fig. 4.— The distribution of pairs observed in the CFRS spectroscopic sample, normalised by the total distribution of pairs expected from all objects with 17.5≤ $I_{AB}$ ≤22.5 in the photometric catalogs. At angular distances <60 arcsec, the number of pairs appears as a linear scaling from all available pairs, while at larger separations the effect of the 3 strip geometry becomes apparent.

Fig. 5.— Number counts of objects for the full CFRS spectroscopic catalog, compared to the number counts from the photometric catalogs. A scaling between the two relations indicates that the total equivalent area surveyed by the CFRS is ∼112 arcmin$^2$, or an effective covering factor of ∼22.4% per field.

Fig. 6.— Example of sky subtraction via polynomial fit of the sky background under the object.

Fig. 7.— Output from MULTIRED for object 10.0496: *(left)* individual corrected 2D spectra and the averaged spectrum; *(right)* 1D spectrum calibrated in wavelength and flux.

Fig. 8.— Raw sky spectrum (top panel) and residual sky fluctuations after sky background removal (bottom panel). The r.m.s. deviation on the residual spectrum is ∼0.25% of the original sky flux.

Fig. 9.— Examples of spectra obtained with CFHT/MOS after processing with MULTIRED, and 3-fold comparison for redshift and confidence class assignment. The object catalog number, $I_{AB}$ magnitude, redshift and confidence class (in brackets), and V-$I_{AB}$ color index are indicated in at the top-left of each spectrum plot. Class (0) is for a non-identified spectrum, class (1), (2), (3) and (4) are for redshift measurements which have 50%, >75%, >95% and 100% chance to be correct. Class (8) is for a single emission line spectrum for which continuum analysis indicates the line is most probably [OII]3727Å, while class (9) is for the cases when the single emission line cannot be identified (see details in CFRS-III). Black squares have been added and represent the photometric mesurements in the V and I bands. This indicates that the spectrophotometric flux calibration of the spectra is better than 10% in most cases.

Fig. 10.— I band images obtained with CFHT/FOCAM (a) the 0000-00 field, (b) the 1000+25 field. The total area represented is 10x10 arcmin$^2$. North is to the top, East to the left.

Fig. 11.— Map of all of the objects with 17.5≤ $I_{AB}$ ≤22.5. The objects with a spectroscopic measurement have been blackened. *(a)* Field 0000-00; *(b)* Field 1000+25.

Fig. 12.— Redshift histograms for galaxies with confidence class 2 and above. (a) field 0000-00,
(b) field 1000+25.

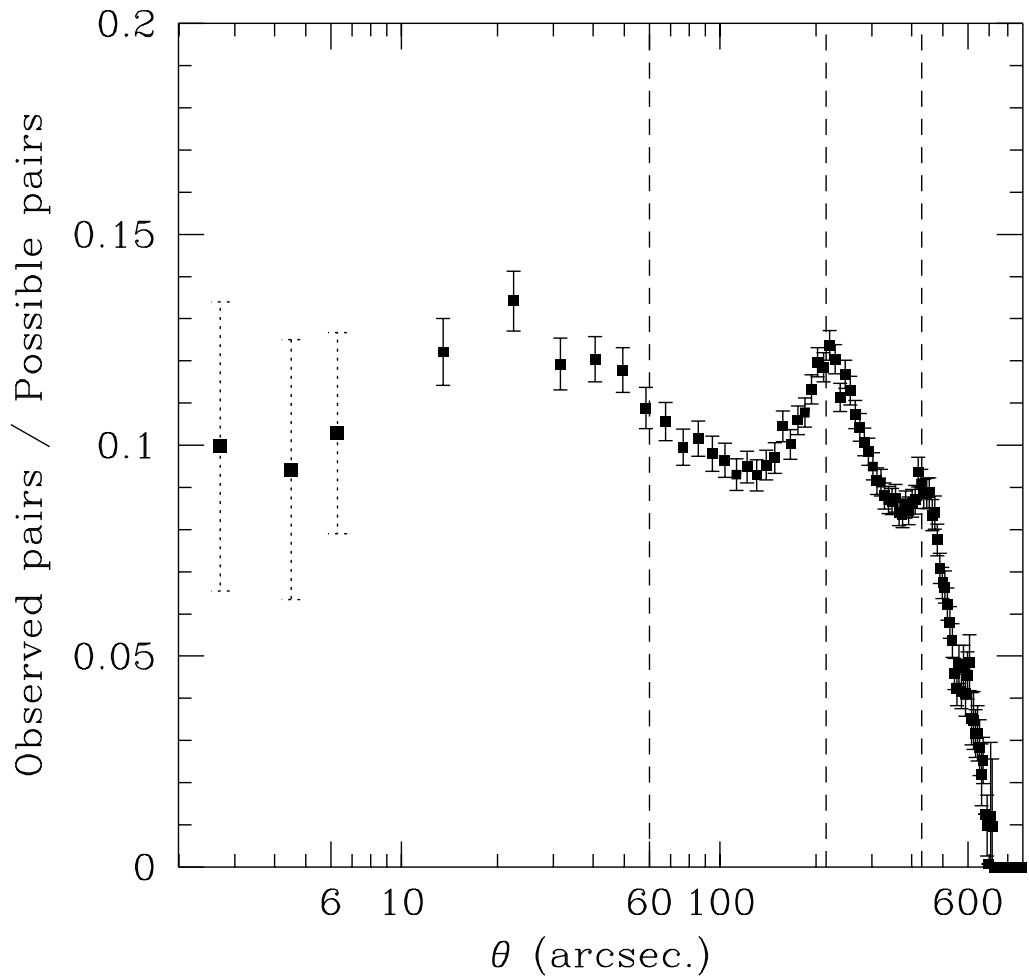

Fig.4

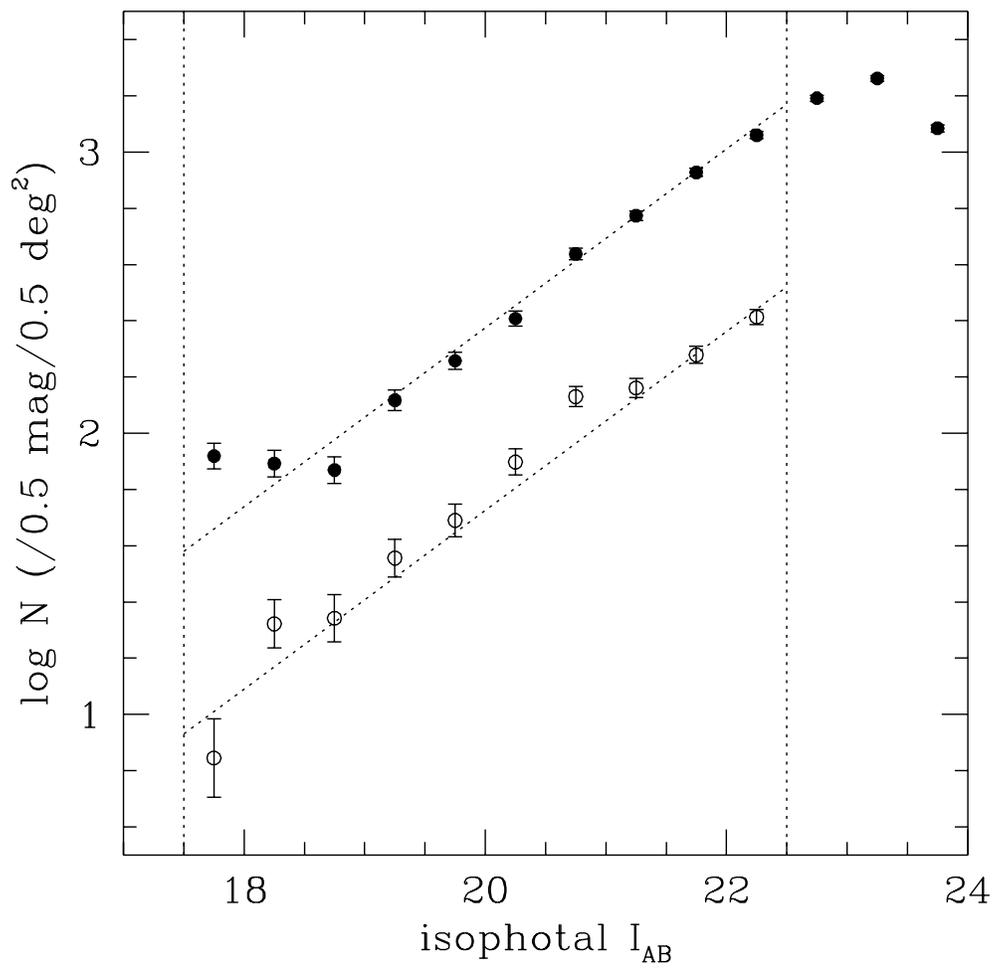

Fig.5

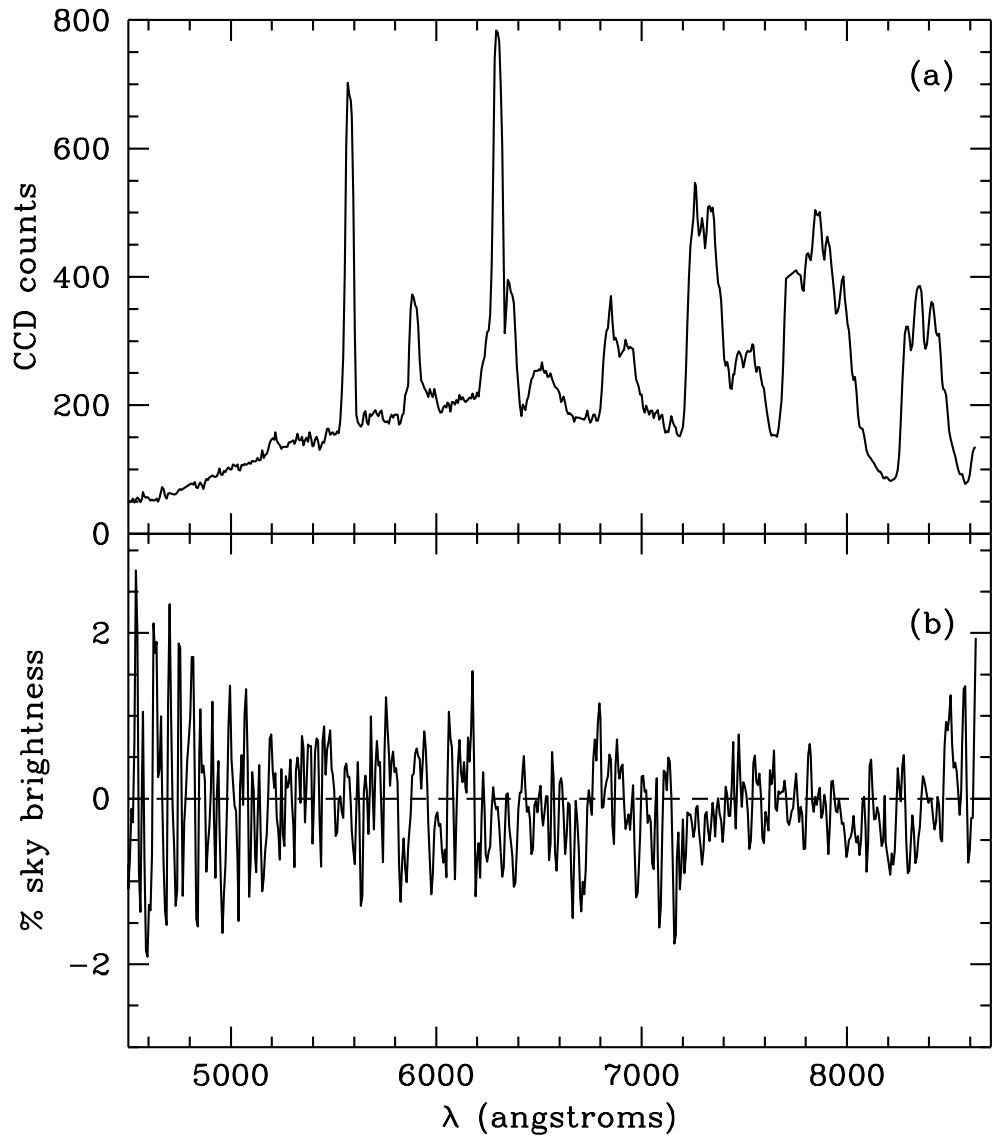

Fig.8

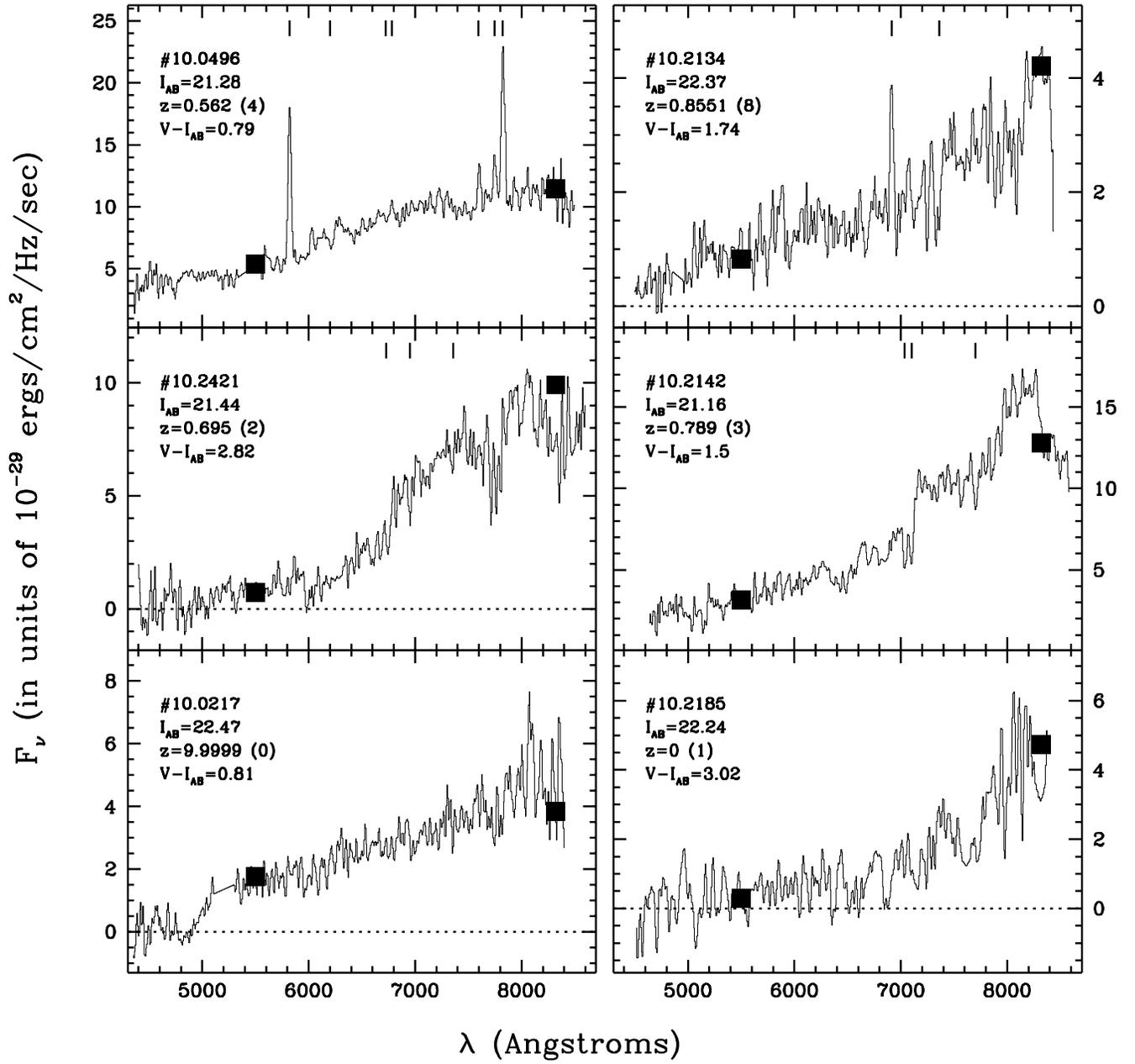

Fig.9

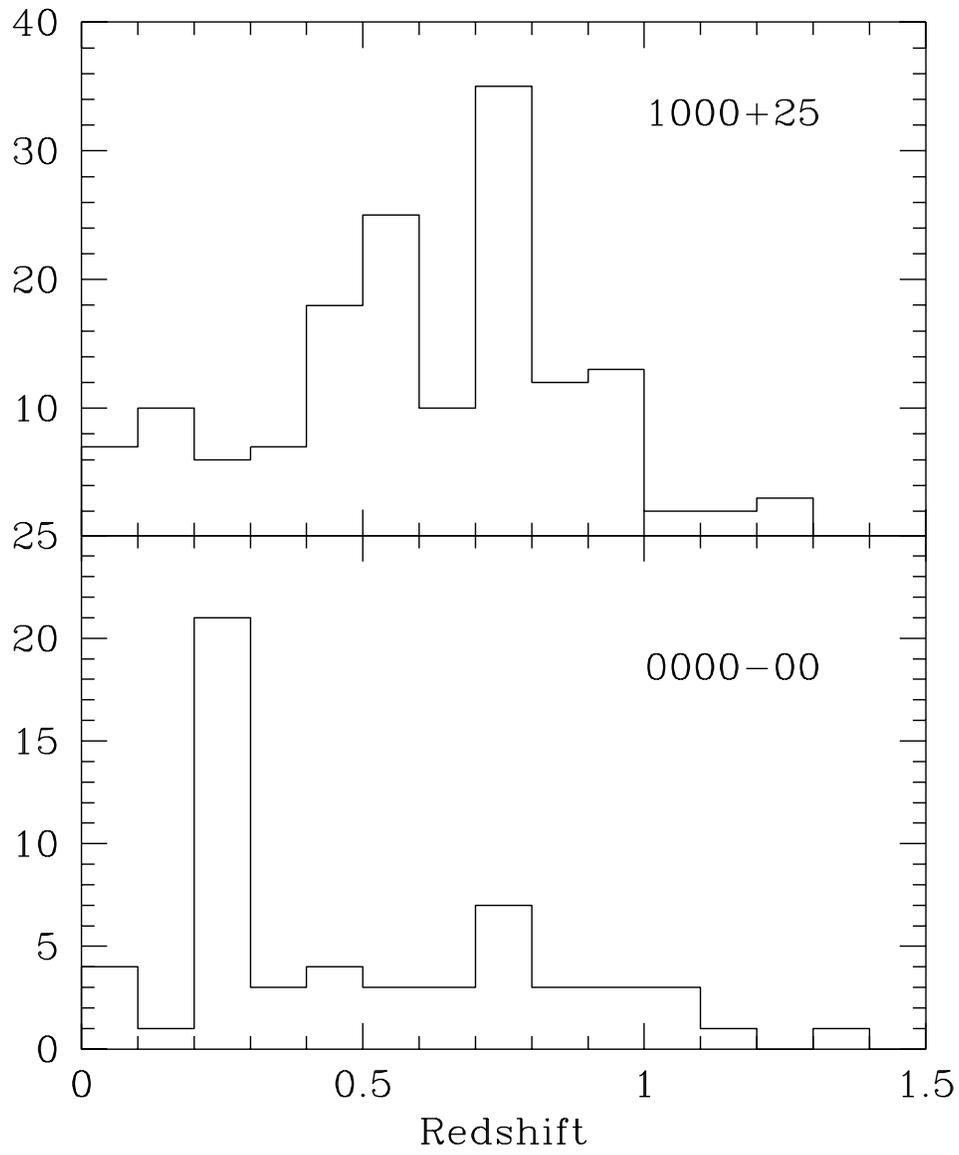

Fig.12